\documentclass{SciPost}
\usepackage{import}

\hypersetup{
    colorlinks,
    linkcolor={red!50!black},
    citecolor={blue!50!black},
    urlcolor={blue!80!black}
}

\usepackage[bitstream-charter]{mathdesign}
\DeclareSymbolFont{usualmathcal}{OMS}{cmsy}{m}{n}
\DeclareSymbolFontAlphabet{\mathcal}{usualmathcal}

\fancypagestyle{SPstyle}{
\fancyhf{}
\lhead{\colorbox{scipostblue}{\bf \color{white} ~SciPost Physics }}
\rhead{{\bf \color{scipostdeepblue} ~Submission }}

\fancyfoot[C]{\textbf{\thepage}}
}

\begin{document}

\pagestyle{SPstyle}

\begin{center}{\Large \textbf{\color{scipostdeepblue}{
Instability of explicit time integration for strongly quenched dynamics with neural quantum states\\
}}}\end{center}

\begin{center}\textbf{
Hrvoje Vrcan\textsuperscript{1$\star$} and
Johan H. Mentink\textsuperscript{1$\dagger$}
}\end{center}

\begin{center}
{\bf 1} Radboud University, Institute of Molecules and Materials, \\
Heyendaalseweg 135, 6525AJ Nijmegen, The Netherlands
\\[\baselineskip]
$\star$ \href{mailto:hrvoje.vrcan@ru.nl}{\small hrvoje.vrcan@ru.nl}\,,\quad
$\dagger$ \href{mailto:johan.mentink@science.ru.nl}{\small johan.mentink@ru.nl}
\end{center}

\togglefalse{showboxes}
\togglefalse{bluecomments}
\togglefalse{redcrossed}

\section*{\color{scipostdeepblue}{Abstract}}
\textbf{\boldmath{%
Neural quantum states have recently demonstrated significant potential for simulating quantum dynamics beyond the capabilities of existing variational ans\"{a}tze. However, studying strongly driven quantum dynamics with neural networks has proven challenging so far. Here, we focus on assessing several sources of numerical instabilities that can appear in the simulation of quantum dynamics based on the time-dependent variational principle (TDVP) with the computationally efficient explicit time integration scheme. \bluecomment{Focusing on}the restricted Boltzmann machine architecture, we compare solutions obtained by TDVP with analytical solutions and implicit methods as a function of the quench strength. Interestingly, we uncover a quenching strength that leads to a numerical breakdown in the absence of Monte Carlo noise, despite the fact that physical observables don't exhibit irregularities. This breakdown phenomenon appears consistently across several different TDVP formulations, even those that eliminate small eigenvalues of the Fisher matrix or use geometric properties to recast the equation of motion. \bluecomment{We provide evidence that the nature of the instability stems from stiffness of the dynamics of the variational parameters, despite the absence of stiffness in the exact quantum dynamics.}We conclude that alternative methods need to be developed to leverage the computational efficiency of explicit time integration of the TDVP equations for simulating strongly nonequilibrium quantum dynamics with neural-network quantum states.
}}

\vspace{\baselineskip}

\noindent\textcolor{white!90!black}{%
\fbox{\parbox{0.975\linewidth}{%
\textcolor{white!40!black}{\begin{tabular}{lr}%
  \begin{minipage}{0.6\textwidth}%
    {\small Copyright attribution to authors. \newline
    This work is a submission to SciPost Physics. \newline
    License information to appear upon publication. \newline
    Publication information to appear upon publication.}
  \end{minipage} & \begin{minipage}{0.4\textwidth}
    {\small Received Date \newline Accepted Date \newline Published Date}%
  \end{minipage}
\end{tabular}}
}}
}


\vspace{10pt}
\noindent\rule{\textwidth}{1pt}
\tableofcontents
\noindent\rule{\textwidth}{1pt}
\vspace{10pt}



\section{Introduction} \label{sec:introduction}
Access to quantum dynamics of many-body systems is of key relevance to many research fields in physics and chemistry, and in particular, in condensed matter \cite{eisert_quantum_2015}. For problems that usually appear in these fields, accessing large quantum systems is key to a more complete understanding of quantum phenomena. However, access to large quantum systems is rendered notoriously difficult by the exponential growth of the Hilbert space \cite{lin_exact_1993, wu_variational_2023}. Recently, Neural Quantum States (NQS) have emerged as powerful methods that bypass the limitations imposed by existing methods \cite{lu_universal_2020, rzadkowski_artificial_2022, lange_architectures_2024}, showing fascinating results for some of the most challenging 2D systems \cite{carleo_solving_2017, schmitt_quantum_2020, fabiani_supermagnonic_2021, zhang_paths_2025}. However, the results reported so far are limited to dynamics near the ground state, and strongly driven quantum dynamics are generally considered highly challenging \cite{hofmann_role_2022, czischek_quenches_2018, gravina_neural_2024, gutierrez_real_2022, lin_scaling_2022}.

A standard approach to obtaining dynamics from a variational representation is the Time-Dependent Variational Principle (TDVP) \cite{kramer_geometry_1981, kramer_review_2008, haegeman_timedependent_2011, ido_timedependent_2015}, first formulated by Dirac. TDVP generates a system of nonlinear ordinary differential equations (ODE) of motion for the variational parameters, which can be solved with standard explicit integration schemes \cite{hjorth-jensen_computational_2010}, providing access to large systems \cite{carleo_solving_2017, fabiani_supermagnonic_2021, zhang_paths_2025, schmitt_quantum_2020}. However, this ODE system requires Monte Carlo sampling of the energy and wave function gradients \cite{becca_quantum_2017} entering the equations, making it prone to stochastic noise. Typically, the combination of noise and nonlinearities leads to a buildup of numerical errors \cite{hofmann_role_2022, donatella_dynamics_2023, walle_manybody_2024}, eventually destabilizing the integration. Stochastic sampling of states with a small contribution to the wave function can also lead to wrong estimations of parameter updates \cite{sinibaldi_unbiasing_2023}. In addition, the variational representation of the quantum wave function is singular in almost all cases, which makes TDVP equations mathematically ill-defined \cite{hackl_geometry_2020, fabiani_quantum_2022}. A standard trick for this issue is to introduce a regulator: a mathematical artefact that can also lead to numerical instabilities \cite{king_computational_2024, zhang_paths_2025, carleo_solving_2017, hofmann_role_2022, gravina_neural_2024}. However, it remains unclear if the above-mentioned sources of instabilities are necessary and sufficient to explain the reported numerical breakdowns and truly limit simulating strongly driven quantum dynamics.

Recently, several works have moved towards improving the NQS time evolution by exploring alternatives to the standard TDVP. \bluecomment{For example, in contrast to solving a continuous-time ODE of variational parameters, methods introduced in \cite{gutierrez_real_2022, zhang_paths_2025, gravina_neural_2024, donatella_dynamics_2023, sinibaldi_unbiasing_2023} solve an optimization problem at each step. In \cite{gutierrez_real_2022}, the time evolution is obtained with the implicit midpoint integrator, while optimizing the error between the variational and exact propagation at each time step. In \cite{sinibaldi_unbiasing_2023}, the overlap between variational and exact evolution is directly optimized at each time with Suzuki-Trotter decomposition, which is then used to calculate the forward-propagation gradient of variational parameters.}In \cite{walle_manybody_2024, sinibaldi_timedependent_2024}, forward-propagation and integration are replaced with learning the entire quantum time evolution globally. Yet another approach utilizes the autoregressive property of neural networks to obtain stable time evolution through normalized wave function representations \cite{donatella_dynamics_2023}. However, compared to the standard realization of TDVP, these approaches require significantly more computational efforts, while the actual dynamics might not even require such complexity. Furthermore, the use of these advanced methods still does not guarantee that all sources of numerical instabilities are addressed, even for simple dynamical scenarios.

Therefore, in this paper, we approach the problem of numerical instabilities differently. We seek to identify the origin of the problematic part of the TDVP time integration, rather than replace it with a more complex method. Attempts at this could already be drawn from the literature. For example, \cite{kelly_adaptive_2022} introduced a taming scheme that can be used to rescale the energy gradient in TDVP, which could stabilize the integration if the gradient becomes too large. Additionally, several reformulations of TDVP have been introduced, like working in the regular subspace of the diagonalized Fisher matrix \cite{schmitt_quantum_2020, dash_efficiency_2025}, or using the geometry of the variational manifold to recast TDVP equations \cite{hackl_geometry_2020}. However, a systematic comparison between these improvements to the standard method is missing. Similarly, assessing the importance of the mentioned sources of instabilities and potentially identifying new ones is missing as well. Thus, we critically assess the possible sources of inaccuracy that can appear in TDVP using a numerically cheap explicit time integration method. As a benchmark, we compare these results with exact diagonalization (ED) and implicit time integration. In all cases, we remove the sampling noise from the TDVP equation of motion by calculating the quantum averages over the whole Hilbert space. Furthermore, we study the significance of regularization by comparing it with two additional formulations of TDVP designed to make the integration regular. Finally, to assess the impact of singularity on time integration, we observe the spectrum of the quantum geometric tensor.

This paper is organized as follows. In Sec. \ref{sec:methods}, we describe the Hamiltonian and the neural network models, as well as all the TDVP formulations and integrators. In Sec \ref{sec:results}, we present the correlation dynamics obtained by integrating the TDVP equation of motion for various driving amplitudes and with various methods, unveiling an undocumented numerical breakdown regime. In Sec. \ref{sec:discussion}, we discuss the results and their implications. Finally, we conclude and provide an outlook in Sec. \ref{sec:conclusion}.

\section{Methods} \label{sec:methods}
In this work, we study a system of antiferromagnetically coupled quantum spins on a 2D lattice, interacting according to the nearest-neighbour Heisenberg model:
\begin{equation}
    H = J_0 \sum_{\{i,j\} \in X,Y} \mathbf{S}_i \cdot \mathbf{S}_j + J(t) \sum_{\{i,j\} \in Y} \mathbf{S}_i \cdot \mathbf{S}_j,
    \label{eq:heisenberg}
\end{equation}
where $\mathbf{S}_i$ is the spin operator on the $i$-th site. The sum over $\{i,j\} \in X,Y$ is taken over all nearest-neighbour pairs of the lattice, while $\{i,j\} \in Y$ indicates perturbation of vertical bonds by the function $J(t)$. This is a minimal model to represent the terahertz dynamics of magnetic systems driven by an optical perturbation of exchange interactions \cite{zhao_magnon_2004, bossini_macrospin_2016, bossini_laserdriven_2019, formisano_coherent_2024}. The system is prepared in the ground state and driven by a quench-like perturbation:
\begin{equation}
    \label{eq:quench}
    J(t) = \begin{cases}
        0, & t<0 \\
        \Delta J_0, & t\geq 0
    \end{cases}.
\end{equation}

As a variational representation, we use the archetypical Restricted Boltzmann machine \cite{carleo_solving_2017}:
\begin{equation}
    \Psi (s) = \prod_{j=1}^M 2\cosh{(\theta_j (s))}.
    \label{eq:rbm}
\end{equation}
Here, $s=\{s^z_i\},\: i=1,\dots, N$ is the spin configuration of $N$ particles, $M=\alpha N$ determines the expressive power of the network parametrized by $\alpha$, and $\theta_j = b_j + \sum_i s_i^z w_{ij}$ includes the biases $b_j$ and weights $w_{ij}$ of the network. This network has been successful in representing a wide variety of quantum spin models \cite{carleo_solving_2017, lange_architectures_2024}, leveraging physical symmetries to reduce the number of network parameters \cite{carleo_solving_2017, choo_symmetries_2018, carrasquilla_machine_2020, fabiani_investigating_2019}, providing access to large systems. Examples include the ultrafast dynamics in the antiferromagnetic Heisenberg model \cite{fabiani_investigating_2019, fabiani_supermagnonic_2021, fabiani_ultrafast_2022, fabiani_quantum_2022, hofmann_role_2022}, and the transverse-field Ising model \cite{zhang_paths_2025, schmitt_quantum_2020, carleo_solving_2017}. The time dependence of the neural network is encoded in the time dependence of its parameters. These follow the TDVP equation of motion \cite{kramer_geometry_1981, kramer_review_2008}:
\begin{equation}
    \label{eq:tdvp}
    S_{kk'} \dot{W}_{k'} = -iF_k,
\end{equation}
where elements $W_{k'}$ include all the RBM parameters. The elements $F_k = \left< E_{\mathrm{loc}} O^*_k\right>- \left< E_{\mathrm{loc}}\right>\left< O^*_k\right>$ constitute the energy gradient vector in the parameter space. The covariance matrix elements $S_{kk'} = \left< O^*_k O_{k'}\right>- \left< O^*_k\right>\left< O_{k'}\right>$ define the quantum Fisher matrix (QFM) \cite{fabiani_quantum_2022}, which is the metric of the parameter space of the selected network \cite{king_computational_2024, hackl_geometry_2020}. Here, $\left<\cdot \right>$ represents the quantum-mechanical average over the entire Hilbert space. The logarithmic derivative functions are defined as $O_k (s) =1/\Psi(s) \cdot \partial_{W_k} \Psi(s)$, while the local value of energy is $E_{\mathrm{loc}}(s) = \left<s\right|\hat{H}\left|\Psi\right>/ \Psi(s)$. 

The TDVP equation consists of a set of first-order differential equations for network parameters. Since these equations are nonlinear, even for the simplest neural network architectures, numerical integration is unavoidable to solve them. For this task, we consider three different formulations of TDVP, which we refer to as \textit{regularization} \cite{becca_quantum_2017, wu_variational_2023}, \textit{diagonalization} \cite{schmitt_quantum_2020, dash_efficiency_2025}, and the \textit{geometric method} \cite{hackl_geometry_2020}. We use these formulations to solve the TDVP equation of motion with an explicit integrator, and also compare this with implicit integration. To describe what these methods do, we rewrite the TDVP equation of motion as $\dot{\mathbf{W}} = \mathbf{f}(\mathbf{W})$, where a vector is defined in bold as $\mathbf{W} = (W_1, W_2, \dots)$. Here, $\mathbf{W} = \mathbf{W}(t)$ is the vector of all network parameters, and the update function $\mathbf{f}$ is obtained by solving Eq. \eqref{eq:tdvp} at some time $t$. Then, an integrator defines a way to calculate the parameter vector in the next integration step $\mathbf{W}_{p+1} = \mathbf{W}(t+\mathrm{d}t)$, using the current $\mathbf{W}_p = \mathbf{W}(t)$. A formulation defines a way to obtain the update $\mathbf{f}$. Formally, the update function is the inverse of Eq. \eqref{eq:tdvp}:
\begin{equation}
\label{eq:update}
 f_k  = -i S^{-1}_{kk'}F_{k'} .
\end{equation}
The $S$-matrix is singular in general, and therefore non-invertible, which means that Eq. \eqref{eq:update} denotes a Penrose-Moore pseudoinverse \cite{penrose_generalized_1955}. 
\begin{table}[ht]
\caption{Integrators and formulations used to solve the TDVP equation of motion \eqref{eq:tdvp}. A more detailed overview of formulations can be found in Appendix \ref{sec:appendix_formulations}. In regularization, $\mathds{1}$ is a unit matrix. For diagonalization, $\mathrm{Diag}(S)$ denotes the $S$-matrix in its diagonal basis, and subscripts $\mathrm{zero}$ and $\mathrm{nonzero}$ denote the singular and the nonsingular subspaces of the basis, respectively. In the geometric method,  $\mathbf{f}_{\mathrm{geo}}$ and $\mathbf{\lambda}$ form a vector which solves Eq. \eqref{eq:tdvp_geo}, and the Lagrange multipliers $\mathbf{\lambda}$ are discarded.} \label{tab:methods}
\centering
\renewcommand{\arraystretch}{1.6}
\makebox[\textwidth][c]{
\begin{tabular}{p{0.3\textwidth}|p{0.675\textwidth}}
\hline
\rowcolor[gray]{0.95}
\multicolumn{2}{c}{
  \textbf{Integrators}
} \\
\hline \hline
\raggedleft Explicit Heun's scheme &
{\rule{0pt}{5ex}\raggedleft\begin{minipage}{\linewidth}
\begin{equation}
    \label{eq:heun}
    \mathbf{W}_{p+1} = \mathbf{W}_p + \dfrac{\mathrm{d}t}{2} \bigg( \mathbf{f}( \mathbf{W}_p ) + \mathbf{f}\left( \mathbf{W}_p + \mathrm{d}t\mathbf{f}( \mathbf{W}_p)\right) \bigg)
\end{equation}
\end{minipage}}
\vspace{0.5pt}
\\
\hline
\raggedleft Implicit midpoint &
{\rule{0pt}{5ex}\raggedleft\begin{minipage}{\linewidth}
\begin{equation}
    \label{eq:implicit}
    \mathbf{W}_{p+1} = \mathbf{W}_p + \mathrm{d}t \mathbf{f} \Bigg( \frac{1}{2}\left( \mathbf{W}_p + \mathbf{W}_{p+1} \right)\Bigg)
\end{equation}
\end{minipage}}
\vspace{1pt}
\\
\hline
\rowcolor[gray]{0.95}
\multicolumn{2}{c}{
  \textbf{Formulations}
} \\
\hline \hline
\raggedleft Regularization &
{\rule{0pt}{5ex}\raggedleft\begin{minipage}{\linewidth}
\begin{equation*}
    S\rightarrow S_{\mathrm{reg}} = S +\varepsilon\mathds{1}
\end{equation*}
\begin{equation}
    \label{eq:reg_update}
    \mathbf{f}_\mathrm{reg} = -iS_{\mathrm{reg}}^{-1} \mathbf{F}
\end{equation}
\end{minipage}}
\vspace{0.5pt}
\\
\hline
\raggedleft Diagonalization &
{\rule{0pt}{5ex}\raggedleft\begin{minipage}{\linewidth}
\begin{equation*}
    S \rightarrow \mathrm{Diag}(S) = S_\mathrm{zero} \oplus S_\mathrm{nonzero}
\end{equation*}
\begin{equation*}
    \mathbf{F} \rightarrow \mathbf{F}_\mathrm{zero} \oplus \mathbf{F}_\mathrm{nonzero}
\end{equation*}
\begin{equation}
    \label{eq:dia_update}
    \mathbf{f}_\mathrm{dia} = -i S_\mathrm{nonzero}^{-1} \mathbf{F}_\mathrm{nonzero}
\end{equation}
\end{minipage}}
\vspace{1pt}
\\
\hline
\raggedleft Geometric method &
{\rule{0pt}{5ex}\raggedleft\begin{minipage}{\linewidth}
\begin{equation*}
     S\otimes\begin{pmatrix}1 & i\\-i&1\end{pmatrix} = S_{\mathrm{geo}} \begin{cases} g = \mathrm{Re}S_{\mathrm{geo}} \\ \omega = \mathrm{Im}S_{\mathrm{geo}}\end{cases}, \; \mathbf{F}\otimes \begin{pmatrix}1\\-i\end{pmatrix} = \mathbf{F}_{\mathrm{geo}}
\end{equation*}
\begin{equation}
    \label{eq:tdvp_geo}
    \begin{pmatrix} 2g & \omega^T \\
    \omega & 0 \end{pmatrix}
    \begin{pmatrix} \mathbf{f}_{\mathrm{geo}} \\ \lambda \end{pmatrix}
    = \begin{pmatrix} 0\\ - \mathbf{F}_{\mathrm{geo}} \end{pmatrix}
\end{equation}
\end{minipage}}
\vspace{1pt}
\\
\hline
\end{tabular}
}
\end{table}

An overview of integrators and formulations used in this work is given in Table \ref{tab:methods}. The explicit integrator, and in particular the Heun's scheme \cite{fabiani_quantum_2022, king_computational_2024}, is a standard in TDVP time integration, where $\mathbf{W}_{p+1}$ can be directly calculated from $\mathbf{W}_p$. In contrast, the implicit midpoint update \cite{gutierrez_real_2022} cannot be solved directly, so a root-finding algorithm must be used. We used the Newton-Raphson method \cite{novak_numerical_2022} to solve Eq \eqref{eq:implicit}, implemented in the SciPy package \cite{virtanen_scipy_2020}. Compared to explicit schemes, implicit integration is more accurate, but also more computationally expensive. The three formulations we used each provide a different way to deal with the singularity of the $S$-matrix in Eq. \eqref{eq:update}. Regularization is considered a standard, and it offsets the matrix diagonal by a small value; diagonalization solves the TDVP equation \eqref{eq:tdvp} in the diagonal basis of $S$; the geometric method uses the geometric properties of the variational manifold to recast Eq. \eqref{eq:tdvp} into a linear problem. More details on these formulations can be found in the Appendix \ref{sec:appendix_formulations}. Note that one can use any combination of integrator and formulation to solve the TDVP equation of motion. Finally, we introduce another modification to the TDVP equation of motion: taming, explained in Appendix \ref{sec:appendix_taming}. This procedure rescales the gradient of the update function in Eq. \eqref{eq:update} to control the influence of the nonlinearity of equations of motion in creating numerical instabilities.

Our aim is to first evaluate the accuracy of TDVP integration in the standard scheme, with regularization and explicit integrator, for different quench strengths $\Delta$ in Eq. \eqref{eq:quench}. Next, we explore the possibility of regulator-free time integration by using diagonalization and the geometric method with the explicit integrator. We further assess how explicit integration compares to implicit integration in all formulations, especially for the task of handling numerical instabilities. We track the accuracy and the stability of the time-integration methods by observing: \begin{enumerate*}[label=(\roman*)]
\item the nearest-neighbour correlation function on a quenched bond, and 
\item the spectrum of the S-matrix
\end{enumerate*}. The former captures the leading-order dynamics of observables \cite{formisano_coherent_2024}; thus, we use it to measure accuracy. The latter tells how problematic the $S$-matrix singularity is in performing time integration \cite{dash_efficiency_2025}.

Most of the analysis is carried out on a small $2 \times 2$ lattice. A simple system like this one is small enough to have access to exact diagonalization, which we use as a benchmark for NQS solutions. Contrary to the exact wave function, we obtain the variational wave function by integrating the TDVP equation of motion. Here, we again take advantage of the small system size to calculate quantum averages over the full Hilbert space, which rules out errors due to sampling noise. In addition, we also calculate the dynamics of bigger lattices and network architectures sampled with variational Monte Carlo (VMC) \cite{becca_quantum_2017}, using the ULTRAFAST numerical package \cite{ultrafast-code_ultrafastcode_2024}. The time integration in ULTRAFAST is done with the standard approach: explicit Heun's scheme \eqref{eq:heun} as integrator, and regularization \eqref{eq:reg_update} as formulation. We compare the results of a small system summed over the full Hilbert space to those of bigger, Monte Carlo sampled systems in the same physical setting. This allows us to assess if the same observations apply to bigger systems, wider networks, and stochastic sampling.

In all calculations, the system is initialized in the ground state of the Heisenberg model \eqref{eq:heisenberg}. For NQS variational representations, the ground state is found by a gradient descent algorithm starting from random network parameters. The sign structure of the ground-state wave function is known to obey Marshall's sign rule \cite{marshall_antiferromagnetism_1955}, which can be enforced by a gauge transformation of the Hamiltonian \cite{fabiani_quantum_2022}. The sign rule is obeyed with real network parameters; therefore, we initialize them as real.

\section{Results} \label{sec:results}
In this section, we present the numerical time integration results of an antiferromagnetic $2\times 2$ lattice represented by the RBM neural network. Since our goal is to explore the possibility of computationally efficient time integration of TDVP, we use a simple RBM architecture with only one hidden node. This corresponds to a $\alpha = 1/4$ architecture, which has the same number of parameters as $\alpha=1$ when translational invariance is taken into account. We compare the results of numerical time integration with exact results.

\begin{figure}[ht]
\centering
\includegraphics[scale=0.5]{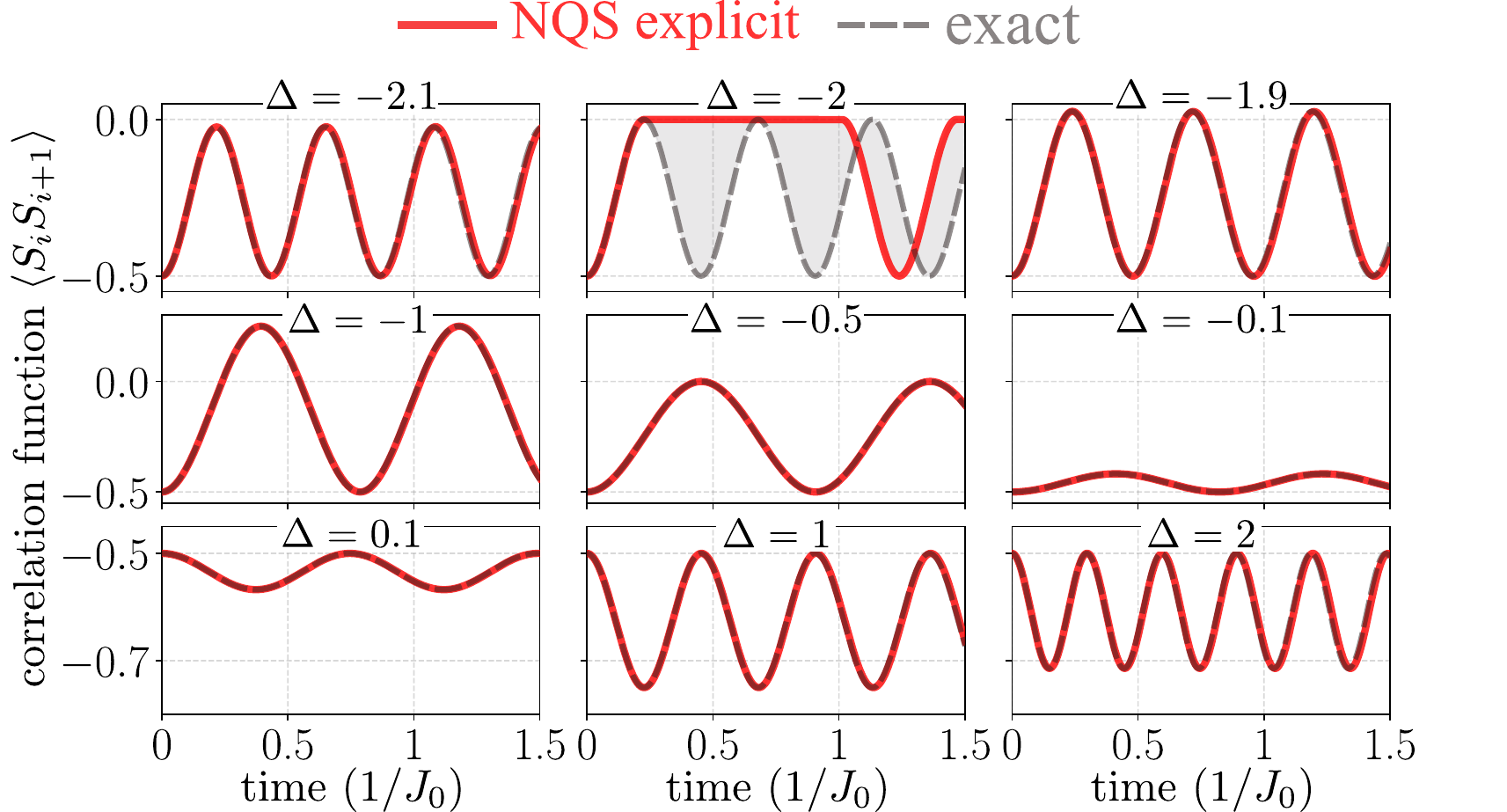}
\caption{Correlation dynamics of NQS compared to ED, as a function of quenching strength $\Delta$. Red lines indicate NQS results, while the dashed gray lines are obtained with ED. Quench strengths are shown on top of each graph. In almost all cases, NQS results agree very well with ED, except for the breakdown quench of $\Delta=-2$.} \label{fig:quenches}
\end{figure}

We first showcase the dynamics of the correlation function $C_{ij} = \braket{S_i \cdot S_j}$ of two spin sites $i$ and $j$ on a quenched vertical bond, as a function of quench parameter $\Delta$ in Eq. \eqref{eq:quench}. To obtain the dynamics, we used a standard approach: the TDVP equation of motion was solved at each time $t$ by regularization formulation Eq. \eqref{eq:reg_update}, and integrated with Heun's update rule Eq. \eqref{eq:heun}. Results of this analysis can be found in Fig. \ref{fig:quenches}. Given the success of the NQS method in various physical scenarios, it is not surprising that the NQS time integration shows an excellent agreement with ED results for almost all values of $\Delta$. This is consistent across various frequencies and amplitudes of correlation oscillations. However, we identify a specific value of $\Delta = -2$, where a numerical breakdown happens. Specifically, when the correlation function reaches the first maximum, all dynamics stop. We refer to this \textit{breakdown point} as a problematic scenario where the explicit TDVP time integration is unable to recover the correct correlation dynamics. 

\begin{figure}[ht]
\centering
\includegraphics[scale=0.4]{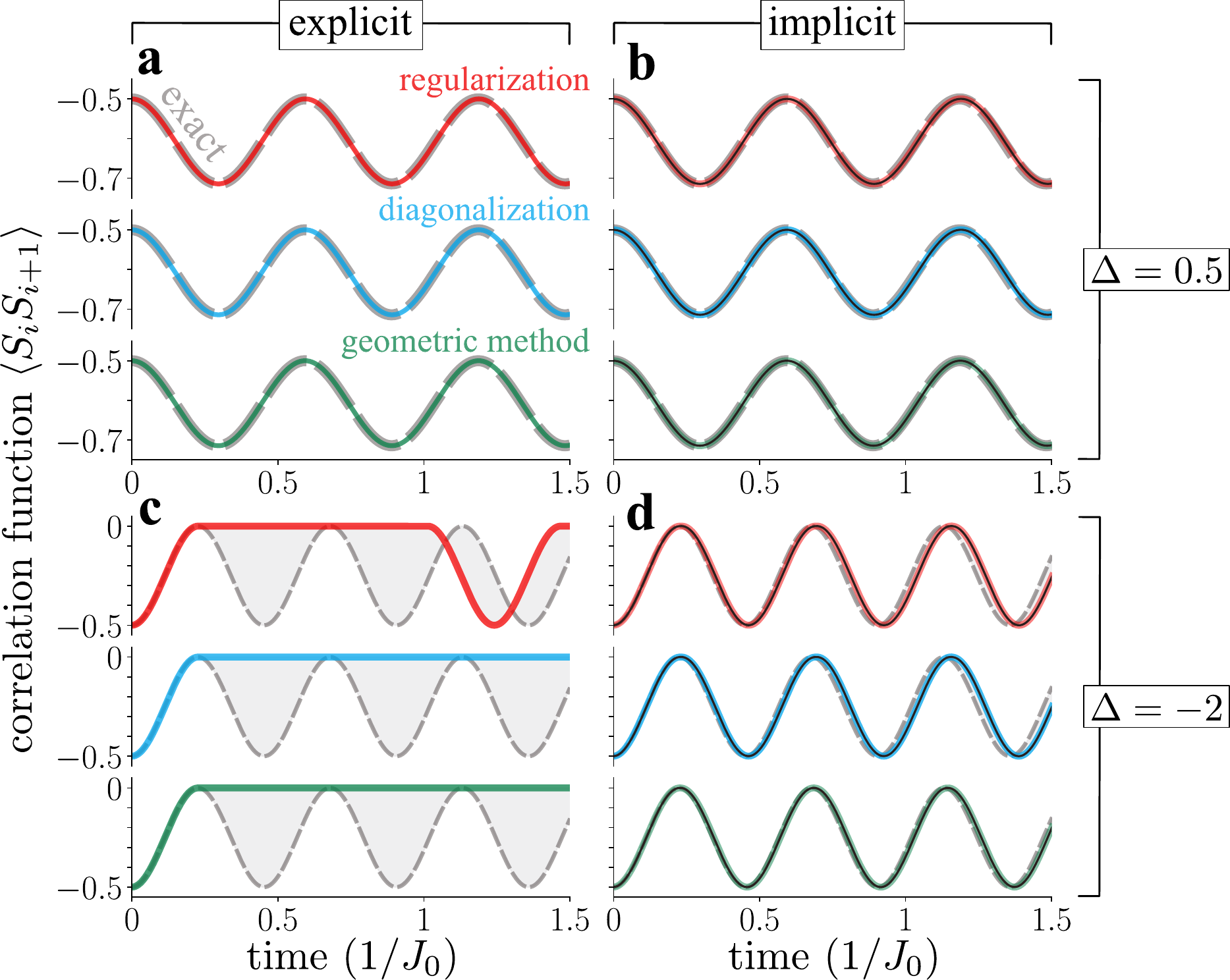}
\caption{Dynamics of correlation function between neighbouring spins along a quenched bond as a function of time, for two quenches: (a, b) top row $\Delta = 0.5$, and (c, d) bottom row $\Delta=-2$. All full lines are obtained by TDVP time integration of NQS represented by the RBM architecture in Eq. \eqref{eq:rbm} with $\alpha=1/4$. Dashed gray lines are the ED results. Colors correspond to different formulations: \textit{regularization} (red), \textit{diagonalization} (blue), \textit{geometric method} (green). All NQS results for the well-behaved $\Delta = 0.5$ agree with the ED results. For the breakdown quench $\Delta=-2$, explicit integrator (left) produces wrong results in all formulations, while the implicit integrator (right) produces correct dynamics. The regularization curve in (c) contains a region of interruption from the frozen dynamics, but still does not recover the correct result.} \label{fig:breakdown}
\end{figure}

As an alternative to the standard approach, we now present results obtained by other formulations of the TDVP equation of motion and the implicit integrator. The results are shown in Fig. \ref{fig:breakdown} for two quenches: $\Delta = 0.5$ and $\Delta = -2$, and for the combination of both integrators and all formulations from Table \ref{tab:methods}. We chose a well-behaved quench strength $\Delta = 0.5$ to demonstrate that accurate integration is possible, even without regularization, and using both integrators. All NQS results in the top row of Fig. \ref{fig:breakdown} follow the exact results. However, in the bottom row, the breakdown quench $\Delta = -2$ shows two qualitatively different behaviours. For explicit integration, the breakdown persists regardless of the formulation used. Therefore, a choice of formulation plays no role in correctly calculating the dynamics for this quench for explicit integration. When we change the integrator to implicit, the accuracy of integration is greatly increased, and correct dynamics are recovered. It should be noted, though, that for $\Delta = -2$, the combination of geometric formulation and implicit integration is the most accurate, as visible by the smallest offset from the exact curve. Thus, for further considerations about the implicit integrator, we used this formulation.

\begin{figure}[ht]
\centering
\includegraphics[scale=0.45]{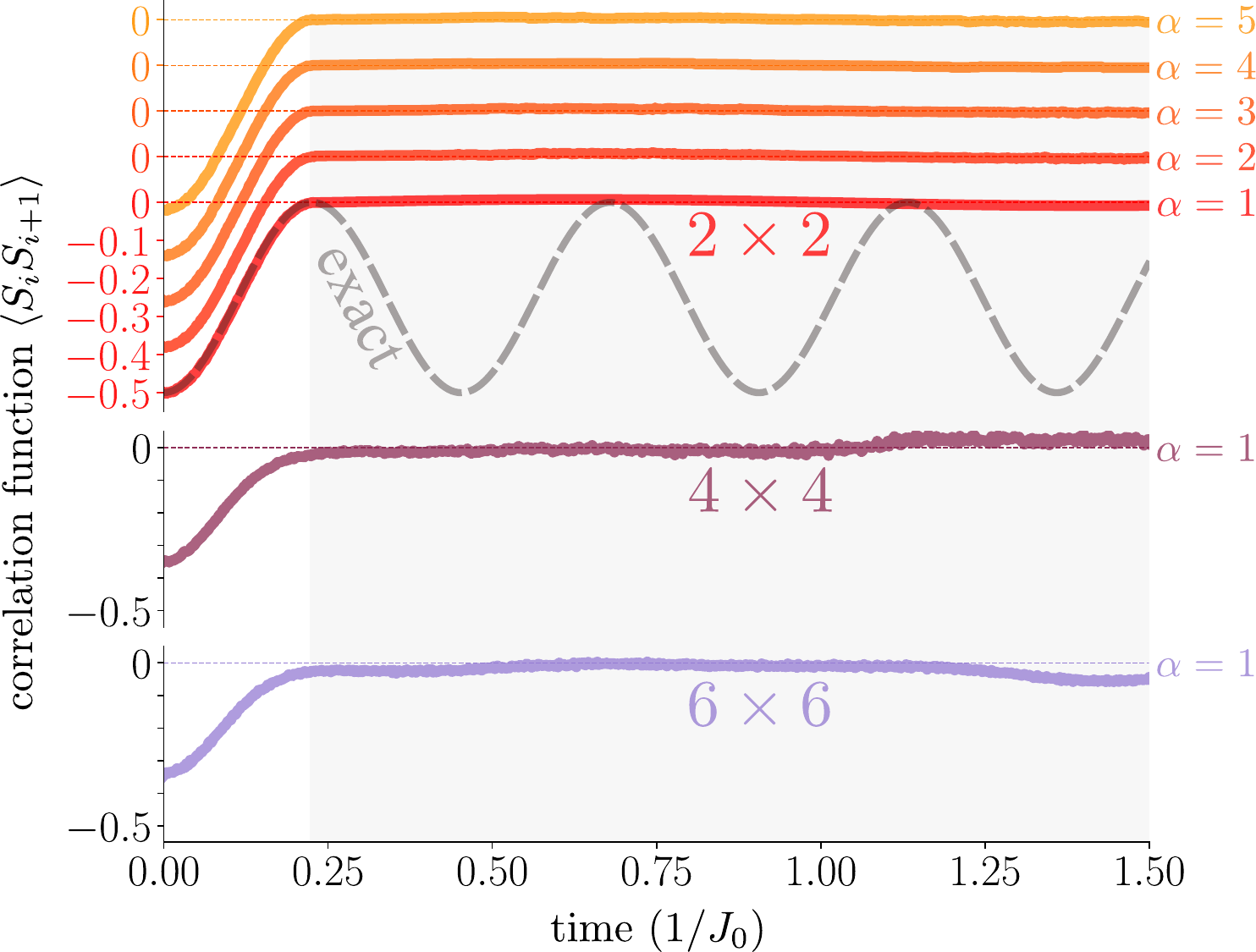}
\caption{Correlation dynamics similar to results in Fig. \ref{fig:breakdown}, for the same value $\Delta = -2$ of quench strength, but for bigger systems and wider neural networks. On the top graph, results are shown for the $2 \times 2$ lattice, and different values of the network width parameter: $\alpha \in \{ 1,2,3,4,5\}$. The dashed grey line refers to the correct result obtained by ED. The two bottom graphs show the same for $4 \times 4$ and $6 \times 6$ lattices, with $\alpha = 1$. All simulations show a numerical breakdown for this quench, characterized by a loss of dynamics after the first maximum of the correlation function (indicated by the shaded area). } \label{fig:ultrafast}
\end{figure}

Next, we present the extension of the breakdown analysis to wider networks and larger lattices. We follow the same recipe as for the $2\times 2$ lattice with $\alpha=1/4$ RBM network architecture. Starting from the ground state of the model, we quench the vertical bonds of the lattice with a $\Delta = -2$ strength, and integrate the TDVP equation of motion with Heun's scheme, in the regularization formulation. These results were obtained with the ULTRAFAST package, where variational Monte Carlo is used to sample the quantum expectation values, unlike the approach shown so far. In Fig. \ref{fig:ultrafast}, we show the dynamics of the correlation function for $\alpha = \{1,2,3,4,5\}$ for the small $2\times 2$ lattice, as well as $4 \times 4$ and $6 \times 6$ lattices with $\alpha = 1$. In all cases, the number of independent parameters is reduced by exploiting the translational symmetry of the lattice. As indicated by the results, the breakdown regime persists across different network architectures and different lattice dimensions, for the same perturbation strength. We also tested bigger network widths $\alpha$ for $4 \times 4$ and $6 \times 6$ lattices, but the results show the same behaviour.

\begin{figure}[h!]
\centering
\noindent\makebox[\textwidth][c]{%
{\large\includegraphics[scale=0.4]{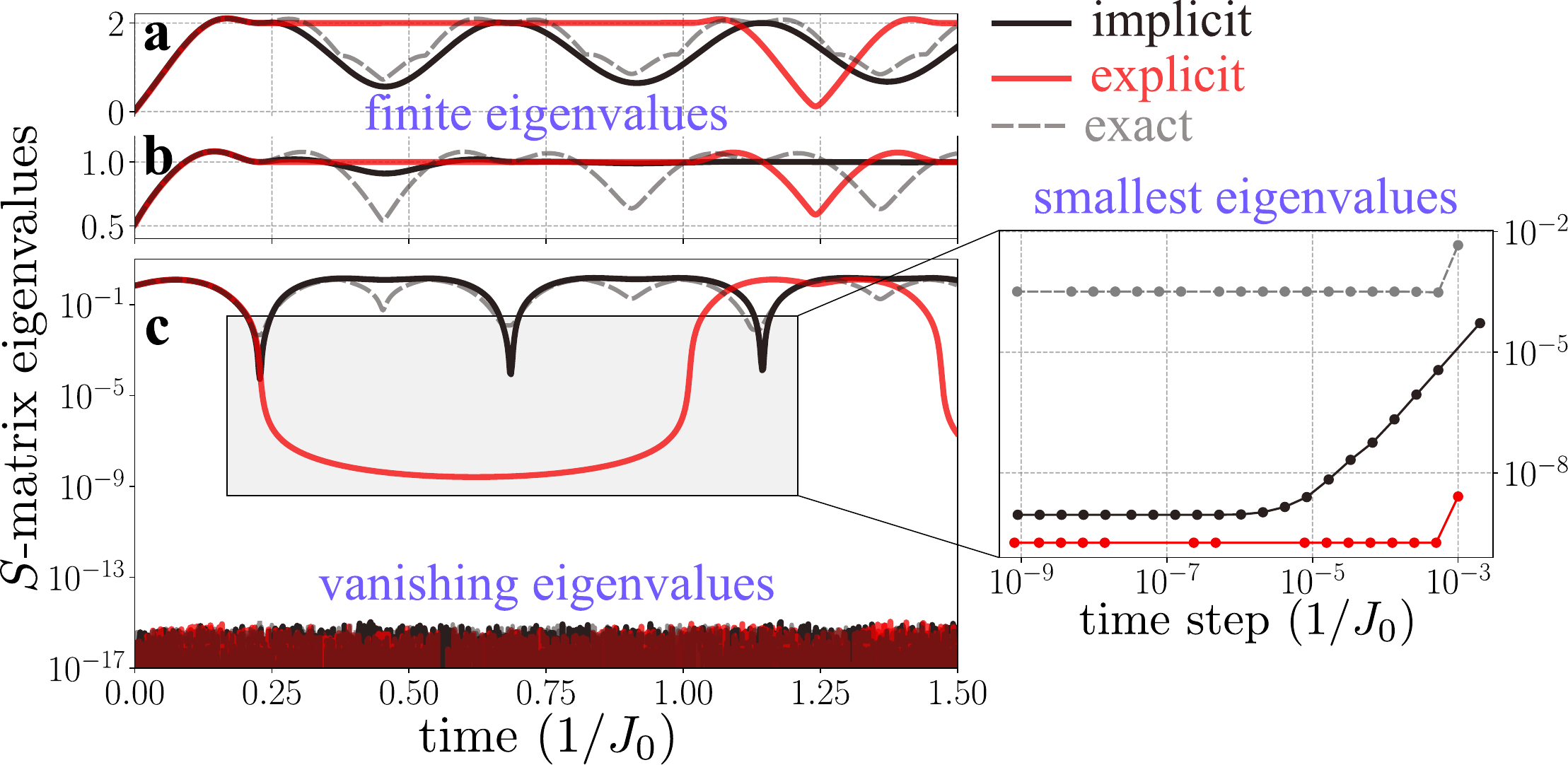}}
}
\caption{Spectra of the Quantum Fisher Matrix Eq. \eqref{eq:tdvp}. The eigenvalues were obtained using three different methods: implicit integration of TDVP using Eq. \eqref{eq:implicit} and the geometric method formulation Eq. \eqref{eq:tdvp_geo}, explicit integration of the TDVP using Eq. \eqref{eq:heun} and the regularization formulation Eq. \eqref{eq:reg_update}, and fitting the RBM architecture to ED solutions using infidelity optimization. The latter is referred to as exact. Three distinct classes of eigenvalues are observed. Finite eigenvalues (a, b) always have a well-behaved value that doesn't cause a singularity. Vanishing eigenvalues (c) always have a value close to zero in machine precision. On the bottom graph, we also see eigenvalues whose value occasionally becomes small for all methods shown. The inset shows the smallest eigenvalues as a function of integration time step, indicating the dependence of eigenvalue cusp depth as a function of numerical accuracy.} \label{fig:spectra}
\end{figure}

To study the effect of singularity on time integration, we observe the spectrum of the $S$-matrix. This matrix is interpreted as a metric tensor of the parameter space \cite{hackl_geometry_2020, king_computational_2024}, so the TDVP equation guides the parameters along a geometrically optimal trajectory. However, if the matrix has zero eigenvalues, there are directions where the evolution of trajectories is unconstrained by the metric. This can lead to numerical instabilities, especially if the trajectory obtains components in these redundant directions. Specifically, we are interested in whether there are new emerging singular directions at the breakdown point. Thus, to assess how the singularity influences the stability of time evolution, we study the spectrum of the $S$-matrix as a function of time. The results are presented in Fig. \ref{fig:spectra} for: 
\begin{enumerate*}[label=(\roman*)]
    \item explicit integration by Heun's scheme,
    \item implicit midpoint integration,
    \item exact solutions obtained by infidelity optimization.
\end{enumerate*} More details calculating the exact RBM representation using infidelity \cite{ledinauskas_universal_2025, donatella_dynamics_2023, dash_efficiency_2025} can be found in Appendix \ref{sec:appendix_ED}. First, we indicate that there are always eigenvalues with values at zero in machine precision, regardless of the integration method, marked as "vanishing eigenvalues". These originate from the overparametrization of the NQS representation. There are also eigenvalues denoted as "finite", whose values are never small throughout the dynamics, so they pose no problem for integration. Secondly, and more interestingly, some eigenvalues occasionally have small values for all the presented methods. These small eigenvalues range from $10^{-9}$ to $10^{-3}$ orders of magnitude, still significantly larger than the machine precision. In particular, the implicit method shows cusps at times coinciding with correlation maxima in Fig. \ref{fig:breakdown} (c. d), the first of which is the breakdown point. The depth of these cusps, or the smallest nonzero eigenvalue, is shown on the inset as a function of integration time step. Decreasing the time step makes the implicit method's cusps reduce to smaller values, saturating around $10^{-9}$. Explicit integration and exact fits are largely unaffected by the reduction of the time step. It should be noted that the dynamics of the correlation function in Fig. \ref{fig:breakdown} are unaffected by the reduction of the time step for all methods presented in this paper.

\section{Discussion} \label{sec:discussion}
The numerical breakdown we introduced in the previous section occurs only at the perturbation parameter $\Delta = -2$. In our analyses, we have not observed an error onset in any quantum observable that would indicate an emergence of the breakdown regime as a function of perturbation. This is visible from the results in the top row of Fig. \ref{fig:quenches}, where two quenches very close in value to the breakdown quench still produce correct dynamics. We stress that all well-behaved results have been obtained by explicit integration in a standard formulation. Furthermore, the NQS numerical error is systematically reducible by reducing the time step for all quenches except at the breakdown. Thus, the origin of the breakdown remains elusive. Even though a quench of this magnitude is physically unrealistic, the cause of the breakdown is not physical, as indicated by the well-behaved ED results. Interestingly, the breakdown is not caused by either stochastic noise or the artificial nature of the regulator. The former is demonstrated by summing over the entire Hilbert space, thus completely removing the sampling noise. The latter is deduced from using two formulations, \textit{diagonalization} and \textit{geometric method}, both of which effectively remove the need for regularization, but still retain the breakdown. 

We emphasize that, even in this regulator-free approach to numerical time integration, with full summation over the Hilbert space, a previously unobserved breakdown regime emerges. This challenges the current understanding of numerical breakdown in variational representations, where instabilities are believed to originate from the interplay between noise, nonlinearity, and singularity. Note that the breakdown could still be caused by the nonlinearity of equations of motion. We introduced taming (Appendinx \ref{sec:appendix_taming}) to deal with this issue, but that might not be sufficient. 

As shown in Fig. \ref{fig:breakdown}, the only way we could treat the breakdown was by using the implicit midpoint integrator. \bluecomment{In Appendix \ref{sec:appendix_comparison}, we show that the parameters of the explicit and implicit integration are very similar up until the breakdown point, after which they become significantly different.}Alongside Heun's \bluecomment{explicit} scheme, we also attempted the Runge-Kutta fourth-order (RK4) scheme, which gave the same results. We also performed an extensive time step analysis for both Heun's and RK4 schemes, which again produced the breakdown for smaller time steps. \crossout{This demonstrates the failure of explicit schemes to capture the correct dynamics in the breakdown scenario, and indicates that explicit schemes with adaptive time steps \cite{schmitt_quantum_2020} still do not recover the correct solution.}On the other hand, implicit schemes work by minimizing the overlap between the left- and right-hand sides of the equation of motion, thus greatly increasing the stability of time integration. However, this minimization comes at great computational costs, as the elements of the TDVP equation of motion have to be evaluated at every trial step. This makes the implicit midpoint method very inconvenient to scale to larger systems.

\begin{figure}[ht]
\centering
\includegraphics[scale=0.5]{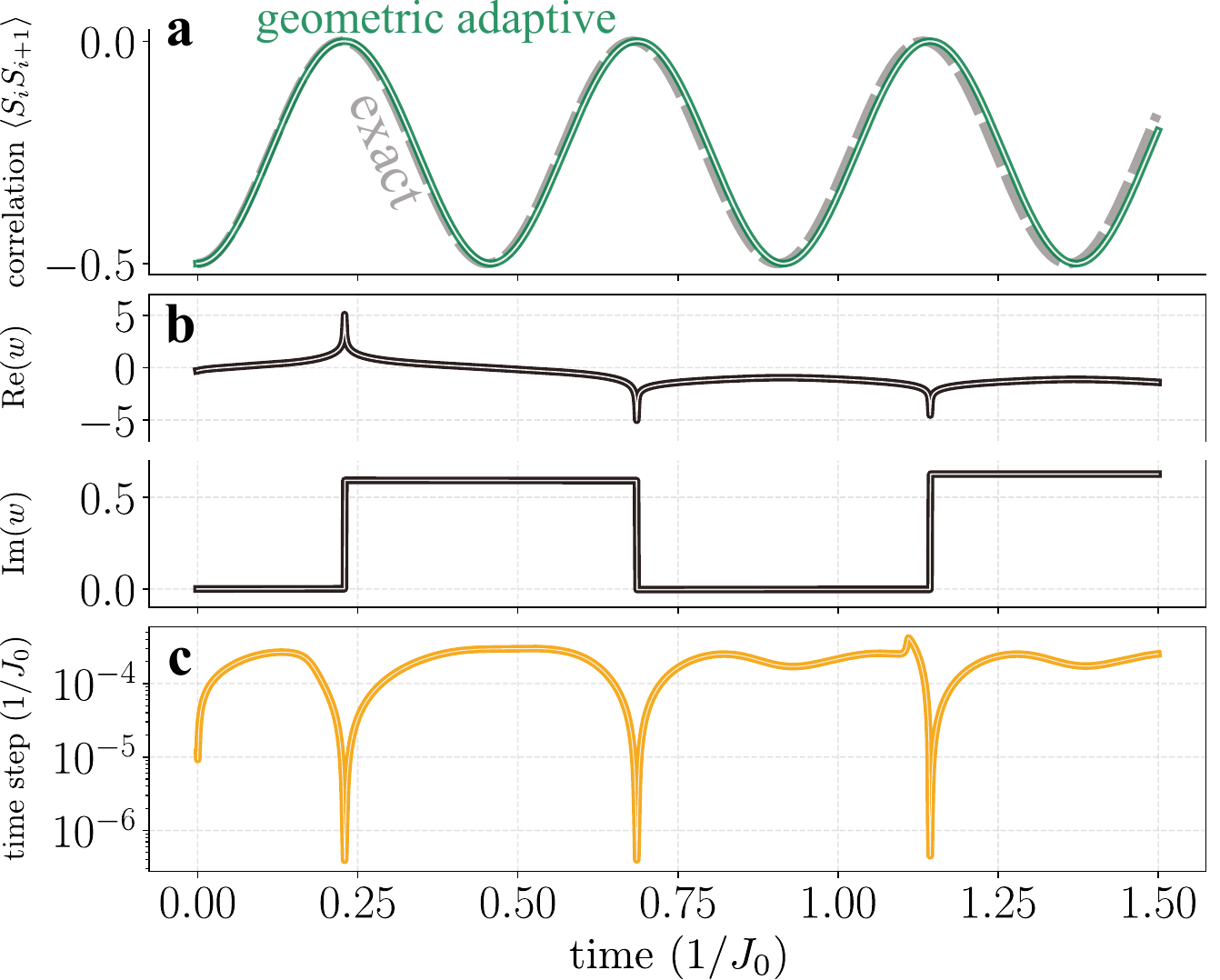}
\caption{Results obtained by the adaptive RK45 integrator. (a) Dynamics of the correlation function, showing excellent agreement with the exact results. (b) Dynamics of real (up) and imaginary (down) components of an example variational parameter $w$, showing cusps at the times of correlation maxima. (c) Adaptive time steps as a function of time. Time steps, initially set to $\mathrm{d}t_0 = 10^{-3}$, reduce drastically at the times of correlation maxima, down to a minimal value $\mathrm{d}t_\mathrm{min} = 3.912\cdot 10^{-7}$.} \label{fig:adaptive_results}
\end{figure}

\bluecomment{We also noticed no drastically varying time scales of dynamics of correlation functions, indicating no physical signatures of stiffness, which could otherwise cause numerical instabilities. However, the dynamics of the variational parameters can still show signatures of stiffness, even if they are not present in the exact quantum dynamics. To test the hypothesis of stiffness being the origin of the instability, we focus on an adaptive integrator. This has been explored before to mitigate instabilities in NQS simulations based on Monte Carlo sampling \cite{schmitt_quantum_2020}. In Appendix \ref{sec:appendix_adaptive}, we provide a formulation of the RK45 adaptive integrator. The result for the breakdown scenario is shown in Fig. \ref{fig:adaptive_results}. We observe that the integrator is capable of recovering the correct dynamics of the correlation function, otherwise problematic for the Heun explicit integrator Eq. \eqref{eq:heun}. However, the variational parameters show cusp-like behavior, as shown for an example parameter in Fig. \ref{fig:adaptive_results} (b), and the adaptive time-stepping leads to a significant reduction of the step size near the breakdown point, as seen in Fig. \ref{fig:adaptive_results} (c). This supports the hypothesis that the nature of the breakdown stems from the stiffness of the variational parameter dynamics. We note that, although the RK45 method is effectively explicit, the large reduction of time step, by more than an order of magnitude, increases the number of steps to reach the same total time, which will be problematic when scaling to large systems. Moreover, calculations of observables with noisy Monte Carlo estimations may put bounds on the tolerance that is feasible for adaptive methods, again leading to instabilities for stiff problems.
}

As far as the scaling properties are concerned, we have shown results for wider networks and bigger lattices in Fig. \ref{fig:ultrafast}. We consistently see the presence of the breakdown regime across all the scaling parameters. Network width is controlled by the parameter $\alpha$ in Eq. \eqref{eq:rbm}, which determines the network's expressive power by setting the number of its hidden neurons. We thus conclude that the breakdown was not caused by an insufficiently expressive network. In fact, the results from implicit integration show that it's possible to represent the correct results, even with low expressivity, with only one RDM hidden node. To further support this claim, we indicate that we were able to fit the network to the correct ED dynamics, as explained in Appendix \ref{sec:appendix_ED}, for the same expressive power. In addition, we observe that the same quenching parameter consistently produces the breakdown across various lattice sizes. Breakdown due to the $\Delta = -2$ quench is therefore not a finite-size effect, but rather a universal phenomenon for bigger lattices as well. \bluecomment{In addition, we tested other neural network models available in the NetKet package \cite{vicentini_netket_2022} on the breakdown scenario (see Appendix \ref{sec:appendix_netket}), which produced instabilities of a similar form. This indicates that the breakdown is not specific to the RBM architecture, but rather a more general problem.}

\bluecomment{
As an additional check, we assessed possible multiscale issues stemming from the exponential form of the wave function. To this end, we performed the calculations of quantum expectation values in two ways:
\begin{enumerate*}[label=(\roman*)]
    \item by using the wave function values ($\Psi$-formulation),
    \item by using the logarithms of the wave function ($\log \Psi$-formulation).
\end{enumerate*} In particular, the latter formulation is useful in dealing with wave function values across multiple orders of magnitude, and is a standard practice in sampling the observables with Monte Carlo methods. As seen in Fig. \ref{fig:psi_logpsi} of Appendix \ref{sec:appendix_logpsi}, we found no difference between the two formulations, showing that the breakdown does not originate from a multiscale problem. Related to different ways to calculate observables, we also considered the breakdown as a consequence of a biasing problem, described in Ref. \cite{sinibaldi_unbiasing_2023}. There, it was pointed out that a \textit{biased} calculation of the energy gradient $F_k$ and the quantum geometric tensor $S_{kk'}$ in Eq. \eqref{eq:tdvp} can lead to wrong trajectories when some elements of the wave function are zero. When using an unbiased formulation, we found that the breakdown is still present. Thus, the way we calculate observables does not influence the breakdown.
}

Finally, we address the singularity of the $S$-matrix in Fig. \ref{fig:spectra}. For all methods, finite eigenvalues don't introduce any difficulties in time integration, and the singularity universally present due to overparametrization can be circumvented. We thus focus on the smallest nonzero eigenvalues, shown in Fig. \ref{fig:spectra} (c). Their values for the exact representation are around $10^{-3}$ order of magnitude, and $10^{-9}$ for the explicit integration. Interestingly, around the breakdown point, the implicit integration yields eigenvalues within the same range, depending on the time step of integration. However, regardless of the time step and therefore the order of magnitude of the smallest nonzero eigenvalue, implicit integration always recovers the dynamics correctly. In contrast, the explicit integration, whose minimal eigenvalue is very close to the saturated value of the implicit case, never recovers the proper dynamics. So, having two cases of very small eigenvalues for two different integrators with distinct accuracies, we conclude that the magnitude of the smallest eigenvalue does not determine the success of time integration. The eigenvalues are still orders of magnitude larger than machine precision, and this is not problematic for numerical calculations in the absence of Monte Carlo noise. Therefore, there is no additional singularity that emerges with time, which would make the explicit integration problematic. Furthermore, the behaviour of small eigenvalues is qualitatively similar for implicit integration and the exact representation. In both methods, eigenvalues have cusps around the times of correlation maxima, while the exact representation has cusps even around the correlation minima, unlike the implicit method. Since these methods both recover the correct dynamics, the magnitudes of the smallest eigenvalues don't seem to be an obstructive factor in the success of time integration. 

\bluecomment{
As a final remark about the singularity, the fact that the $S$-matrix has zero eigenvalues indicates that the TDVP equations of motion are differential-algebraic equations (DAEs), instead of ODEs. In many cases, this necessitates the use of implicit methods to obtain correct results with numerical integration \cite{novak_numerical_2022}. However, all three inversion methods (described under "Formulations" in Table \ref{tab:methods} and in detail in Appendix \ref{sec:appendix_formulations}) are constructed to convert DAE into ODE systems, enabling numerical integration with standard techniques. Thus, the fact that we obtain correct results only with an implicit integrator does not signify that we're dealing with problems of DAE systems, and they are not the cause of the breakdown. Collecting these observations, we conclude that the time integration is possible regardless of the singularity of the $S$-matrix, and the breakdown does not emerge from the singularity. 
}

\section{Conclusion} \label{sec:conclusion}
To conclude, we have presented an analysis of the stability and accuracy of NQS, which represents quenched dynamics of the Heisenberg antiferromagnet with TDVP. We uncovered a numerical breakdown, even with a fully sampled Hilbert space, and without regularization. Interestingly, the breakdown does not originate from known problematic factors in TDVP, showing that numerical time integration with explicit methods is more challenging than anticipated before. \bluecomment{Based on our analysis, stiffness in the dynamics of the variational parameters seems to be the nature of the breakdown, yet standard methods to deal with stiff problems, such as implicit methods and adaptive integrators, are too costly to scale to large systems.} With this paper, we hope to motivate a search for a different approach in obtaining stable and accurate NQS dynamics, free of breakdowns, computationally cheap, and scalable to bigger systems. A path towards this may come from restricting the integration to a more stable part of the variational manifold, such as the normalized subspace \cite{donatella_dynamics_2023}, or reformulating TDVP beyond the current order of expansion \cite{becca_quantum_2017}. \bluecomment{Finally, to deal with numerical instabilities, it may be interesting to develop neural network models in which variational parameters are constrained to avoid stiff dynamics while still maintaining expressibility.}

\section*{Acknowledgements}
This project was funded by the \href{https://www.nimfeia.eu/}{NIMFEIA}: Nonlinear Magnons for Reservoir Computing in Reciprocal Space project of the European Union, under the number \href{https://cordis.europa.eu/project/id/101070290}{101070290}. JHM acknowledges funding by the Dutch Research Council (\href{https://www.nwo.nl/en}{NWO}) via \href{https://www.nwo.nl/en/researchprogrammes/nwo-talent-programme/projects-vidi}{VIDI} project number 223.157 (\href{https://www.nwo.nl/en/projects/vividi223157}{CHASEMAG}). We are grateful to Mikhail Tretyakov and Gabriel Lord for helpful discussions regarding the stability of numerical time integration methods. \bluecomment{We thank Filippo Vicentini and Giuseppe Carleo for useful comments about the biasing problem.}



\begin{appendix}
\numberwithin{equation}{section}

\setcounter{equation}{0}

\section{Infidelity optimization of a variational wave function} \label{sec:appendix_ED}
To obtain the variational representation $\ket{\phi}=\ket{\Psi_{\mathrm{RBM}} (\mathbf{W})}$ of the exact diagonalization wave function $\ket{\psi} = \ket{\Psi_{\mathrm{ED}} (t)}$, we optimize the RBM architecture using infidelity as loss function:
\begin{equation}
    \label{eq:infidelity}
    L \left(\psi, \phi\right) = 1- \dfrac{\braket{\psi|\phi}\braket{\phi|\psi}}{\braket{\psi|\psi}\braket{\phi|\phi}},
\end{equation}
which is a measure of overlap between two vectors. Here, we fix the exact wave function at some time $t$, while varying the parameters $\mathbf{W}$ of the variational ansatz. The update rule for variational parameters is the gradient descent variant:
\begin{equation}
    \label{eq:gradient_descent_infidelity}
    \mathbf{W} (p+1) = \mathbf{W}(p) - \eta S^{-1}\nabla_\mathbf{W} L,
\end{equation}
where $p$ is the optimizaiton step, and $\eta$ the learning rate. Here, $S^{-1}$ is the inverse of the Quantum Fisher matrix, defined in the same way as in Eq. \eqref{eq:tdvp}, and inverted using the regularization procedure described in \ref{sec:regularization}. We calculate the infidelity gradient $\nabla_\mathbf{W} L$ using: 
\begin{equation}
    \label{eq:infidelity_gradient}
    \nabla_{W_k} L = \dfrac{\braket{\Psi_{\mathrm{ED}}|\Psi_{\mathrm{RBM}}}\cdot\left( \braket{O^*_k} \braket{\Psi_{\mathrm{RBM}}|\Psi_{\mathrm{ED}}} - \sum_{s_i} O^*_k(s_i) \Psi_{\mathrm{RBM}}^* (s_i) \Psi_{\mathrm{ED}}(s_i)\right)}{\braket{\Psi_\mathrm{RBM}|\Psi_\mathrm{RBM}}}.
\end{equation}
We've also taken into consideration that $\ket{\Psi_\mathrm{ED}}$ is normalized. In Eq. \eqref{eq:infidelity_gradient}, the logarithmic derivative $O_k(s)$ is the same as described in Eq. \eqref{eq:tdvp}, and $s_i$ is the $i$-th spin configuration in the Hilbert space.

\section{TDVP formulations} \label{sec:appendix_formulations}
In this appendix, we describe how to solve the TDVP equation of motion using three different formulations.
\subsection{Regularization} \label{sec:regularization}
In the most general situation, the $S$-matrix in Eq. \eqref{eq:tdvp} is singular, which prevents the inversion of the equation. The standard approach to deal with this obstacle is regularization: $S\rightarrow S_{\mathrm{reg}} = S +\varepsilon\mathds{1}$, where $\mathds{1}$ is the unit matrix, and $\varepsilon$ is a small number, often in the interval $\left[10^{-5},\: 10^{-3}\right]$. Following the regularization approach, the recipe for finding the elements of the update function $\mathbf{f}(\mathbf{W})$ in Eq. \eqref{eq:update} is:
\begin{equation}
    \label{eq:app_reg_update}
    \mathbf{f}_\mathrm{reg} = -iS_{\mathrm{reg}}^{-1} \mathbf{F}.
\end{equation} 

Introducing this small diagonal offset makes the determinant $\det(S_{\mathrm{reg}})$ finite, thus rendering the inverse $S_{\mathrm{reg}}^{-1}$ well-defined. However, Hofmann et. al. \cite{hofmann_role_2022} demonstrated a fine balance between stability, accuracy, and regularization, which is especially delicate if the numerical method has a stochastic component, like the Monte Carlo method. Improper choice of regularization can lead to numerical instabilities and eventual breakdowns, occurring sooner in dynamics for stronger perturbations. Therefore, due to known problems with regularization, here we formulate regulator-free integration using two alternative approaches for finding the update function in Eq. \eqref{eq:update}.

\subsection{Diagonalization} \label{sec:diagonalization}
This approach follows from the fact that singular matrices have zero eigenvalues, which don't contribute to dynamics. Thus, the first step of the diagonalization method is diagonalizing the $S$-matrix and obtaining the eigenspace transformation matrix $T$. The elements of the TDVP equation of motion are then transformed: $S\rightarrow S_{\mathrm{dia}}=\mathrm{Diag}(S) = T^{-1} S T$, $\mathbf{F}\rightarrow \mathbf{F}_{\mathrm{dia}} = T^{-1} \mathbf{F}$, $\mathbf{W}\rightarrow \mathbf{W}_{\mathrm{dia}} = T^{-1} \mathbf{W}$. The final step is to remove the nullspace obtained from the transformation matrix $T$ by removing elements that correspond to zero eigenvalues, from all terms of the equation. This removal is done numerically, following a predetermined criterion that eigenvalues below a certain value $\zeta$ are considered to be zero. The deletion produces elements $S_\mathrm{nonzero}$ and $\mathbf{F}_\mathrm{nonzero}$ in Eq. \eqref{eq:dia_update}.

The removal of nullspace boils down to just ignoring those elements of $S_{\mathrm{dia}}$, $\mathbf{F}_{\mathrm{dia}}$, and $\mathbf{W}_{\mathrm{dia}}$ that have the indices of zero eigenvalues. This effectively reduces the dimension of the inversion problem. We find that the zero cutoff criterion $\zeta \approx 10^{-12}$ usually works well to leave the $S_\mathrm{nonzero}$ matrix regular. This way, the equation can be inverted and the update function can be calculated without regularization:
\begin{equation}
    \label{eq:app_dia_update}
    \mathbf{f}_\mathrm{dia} = -i S_\mathrm{nonzero}^{-1} \mathbf{F}_\mathrm{nonzero},
\end{equation}
after which, the problem is transformed back into the original parameter basis. Note that if a numerical integrator uses multiple steps, like the Heun update rule Eq. \eqref{eq:heun}, the diagonalization procedure described here is required at each intermediate step.

\subsection{Geometric method}\label{sec:geometric}
This approach is based on the variational methods formulation on a K{\"a}hler manifold, introduced by Hackl et. al. in \cite{hackl_geometry_2020}. The core of the method is to reparametrize the TDVP equation of motion into a problem where all the parameters of the neural network are real by splitting them into real and imaginary components. We thus perform the following transformation:
\begin{equation}
    \label{eq:app_geo_split}
    \mathbf{W} \rightarrow \mathbf{W}_{\mathrm{geo}} = \{ \mathrm{Re}W_1, \mathrm{Im}W_1, \dots, \mathrm{Re}W_M, \mathrm{Im}W_M\},
\end{equation}
which doubles the dimension of the problem. Given this transformation and the properties of the RBM network Eq. \eqref{eq:rbm}, the transformation rule for other elements of the TDVP equation of motion is:
\begin{equation}
    \label{eq:geo_transform}
    S, \mathbf{F} \rightarrow S_{\mathrm{geo}} = S\otimes\begin{pmatrix}1 & i\\-i&1\end{pmatrix}, \; \mathbf{F}_{\mathrm{geo}} = \mathbf{F}\otimes \begin{pmatrix}1\\-i\end{pmatrix}.
\end{equation}

After transforming all the elements of the equation, two geometric characteristics of the variational manifold are defined: the metric $g = \mathrm{Re}S_{\mathrm{geo}}$, and the symplectic form $\omega = \mathrm{Im}S_{\mathrm{geo}}$. These are used to define the new TDVP equation of motion. In addition, following the prescription in \cite{hackl_geometry_2020}, we can recast the pseudoinversion problem required to solve Eq. \eqref{eq:tdvp} into:
\begin{equation}
    \label{eq:app_tdvp_geo}
    \underbrace{ \begin{pmatrix} 2g & \omega^T \\
    \omega & 0 \end{pmatrix} } _A
    \underbrace{ \begin{pmatrix} \mathbf{f}_{\mathrm{geo}} \\ \mathbf{\lambda} \end{pmatrix} }_x
    =\underbrace{ \begin{pmatrix} 0\\ - \mathbf{F}_{\mathrm{geo}} \end{pmatrix} }_B,
\end{equation}
where $\lambda$ is a vector of Lagrange multipliers. The multipliers serve to constrain the TDVP equation of motion to a subclass of solutions that have a minimal-length component in the singular subspace of the matrix $\omega$. This constraint is meant to eliminate potential instabilities caused by the singularity of the matrix.

This is now a linear problem in the form $Ax = B$ whose dimension is four times the dimension of the original formulation. However, even though the network redundancies that cause the singularity of the $S$-matrix are now recast, the problem still contains them in the matrix $A$. Therefore, to find a solution to this linear problem, we must use a convergence method which minimizes the distance between the left- and right-hand sides of the equation \eqref{eq:app_tdvp_geo}. For this task, we used the least-squares algorithm \cite{golub_matrix_2013} implemented in NumPy \cite{harris_array_2020}, which finds the $x$ that minimizes the norm $\left\Vert Ax - B\right\Vert$ in Eq. \eqref{eq:app_tdvp_geo}. We found that the well-known least-squares method performs just as well as some modern algorithms used for this task, such as MINRES and its variants \cite{paige_solution_1975}.

Finally, when the solution is found, we keep only the update function, $\mathbf{f}_{\mathrm{geo}}$, discarding the Lagrange multipliers $\mathbf{\lambda}$.

\section{Taming} \label{sec:appendix_taming}
Taming is a numerical procedure used in explicit integration of differential equations, particularly useful in preventing nonlinearity of equations from causing instabilities \cite{kelly_adaptive_2022}. Its primary purpose is to control the magnitude of the gradient of the equation variables, in situations where the large gradient drives the evolution too far away from the correct trajectory. For a differential equation $\dot{\mathbf{y}} = \mathbf{f}(\mathbf{y},t)$, the gradient is replaces according to the following rule:
\begin{equation}
    \label{eq:taming}
    \mathbf{f} \rightarrow \dfrac{\mathbf{f}}{1+\mathrm{d}t\left\Vert \mathbf{f}\right\Vert},
\end{equation}
where $\mathrm{d}t$ is the numerical integration step.
The taming procedure can be directly applied to the TDVP update function Eq. \eqref{eq:update} in any formulation. We used taming throughout this work, as part of the attempts to control the breakdown for $\Delta = -2$ quench. However, we haven't noticed a benefit of taming in resolving this instability.

 \section{Comparison of variational parameters} \label{sec:appendix_comparison}
\begin{figure}[ht]
\centering
\includegraphics[scale=0.6]{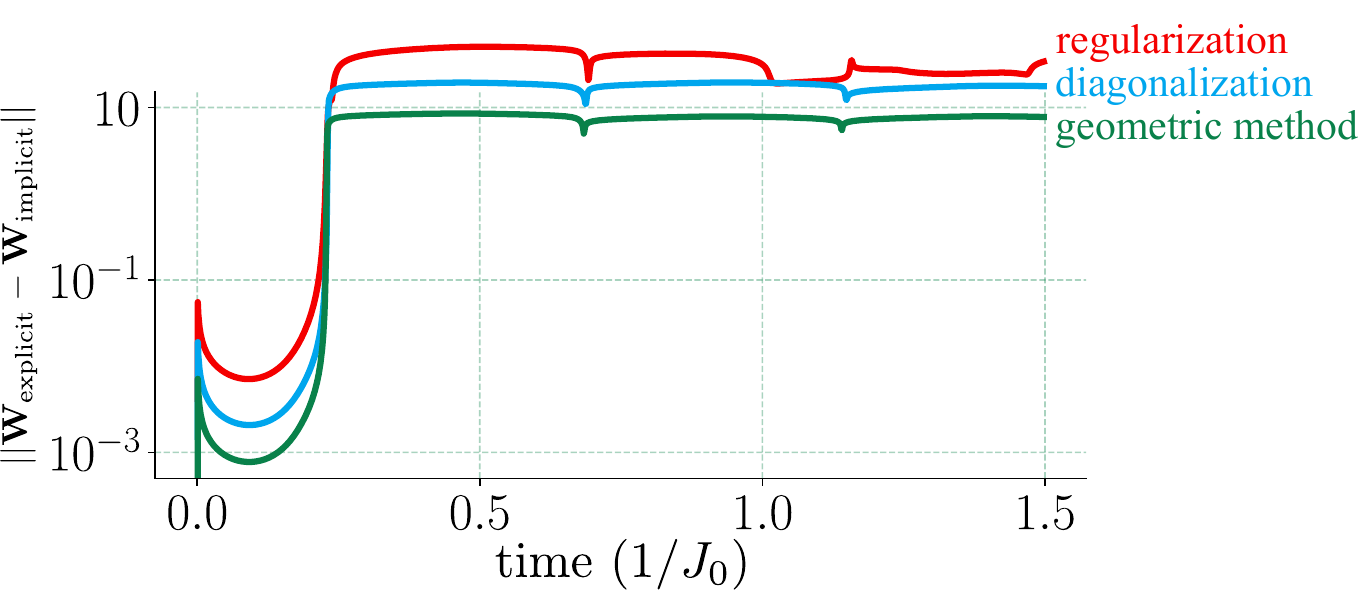}
\caption{Time evolution of the $2$-norm difference between variational parameters for explicit and implicit integrators, for all three formulations used in this work. The grid is shown for the geometric method, while other lines have the same scale, but are offset in the $y$-axis.} \label{fig:weight_differences}
\end{figure}

In this appendix, we compare the differences in the variational parameters between explicit and implicit integration schemes, both starting from the same initial values. We measure the difference between the parameters with a $2$-norm difference:
\begin{equation}
    \left\Vert \mathbf{W}_\mathrm{explicit} - \mathbf{W}_\mathrm{implicit} \right\Vert.
\end{equation}
In Fig \ref{fig:weight_differences}, we observe that both integration schemes yield very similar parameters until the breakdown point. Afterwards, the differences between the parameters become significant, suggesting considerable differences in the variational trajectories following the breakdown. This behavior is consistent across all three formulations used in this work. 

 \section{RK45 adaptive integrator}\label{sec:appendix_adaptive} 
Here, we describe the construction of the Runge-Kutta 4-5 explicit adaptive integrator, and provide some results thereof. The idea behind adaptive integrators is to dynamically modify the time step of numerical time integration to obtain equally accurate results (within a specified tolerance) at each time step. Typically, this is done by predicting the solution to an ODE with two methods of different expansion orders, and then increasing or decreasing the time step based on the difference between those predictions. This technique is useful in handling regions of unstable integration, where reducing the time step can greatly increase the accuracy of the results.

For the two methods of different orders, the Runge-Kutta 4 and Runge-Kutta 5 integrators can be combined into the Runge-Kutta 4-5 (RK45) method. The Butcher tableau of the RK45 integrator we use is given by the Cash-Karp variety \cite{cash_variable_1990}. For the algorithm, we first choose some tolerance factor $\mathrm{tol}$ (for example, $\mathrm{tol} = 10^{-6}$). Then, for an ODE of variable $y$, at time $t$ with time step $h$, we proceed as follows.
\begin{enumerate}
    \item Calculate the predictions of RK4 and RK5 methods from the Butcher tableau, $y_4$ and $y_5$, respectively.
    \item Calculate the error as the norm difference of those predictions: 
    \begin{equation}
            \mathrm{error} = \left\Vert y_5 - y_4\right\Vert,
    \end{equation}
    \item If the error is smaller than the tolerance, $\mathrm{error} < \mathrm{tol}$, update the solution with the higher-level prediction: 
        \begin{equation}
            y(t+h) = y_5.
        \end{equation}
        Here, we also accumulate any observables of the calculation.
    \item Modify the time step according to the formula: 
    \begin{equation}
        h \longrightarrow 0.9\cdot h\cdot \left( \dfrac{\mathrm{tol}}{\mathrm{error}}\right)^{\dfrac{1}{p}},
    \end{equation} 
    where $p$ is the order of the higher-order method, so in the RK45 case, $p=5$.
\end{enumerate}

\section{Other neural network architectures} \label{sec:appendix_netket}
In this appendix, we show results for the $2 \times 2$ Heisenberg antiferromagnet quenched with $\Delta = -2$, using different neural network models as variational ans\"atze. We compared three different models available in the NetKet package \cite{vicentini_netket_2022}: restricted Boltzmann machine (RBMSymm), Jastrow ansatz, and group convolutional neural network (GCNN) \cite{roth_group_2021}. For the RBMSym, we used the $\alpha = 1$ hidden unit density. The Jastrow ansatz architecture only depends on the lattice dimension. For GCNN, we used $1$ hidden layer, with $2$ features in each layer, and the default SELU \cite{klambauer_selfnormalizing_2017} activation function between the layers. In all calculations, the Heun numerical integrator (the same as Eq. \eqref{eq:heun}) was used with the time step of $\mathrm{d}t = 0.001$, and the quantum geometric tensor was regularized with the diagonal shift of $\varepsilon = 10^{-4}$. 

\begin{figure}[ht]
\centering
\includegraphics[scale=0.6]{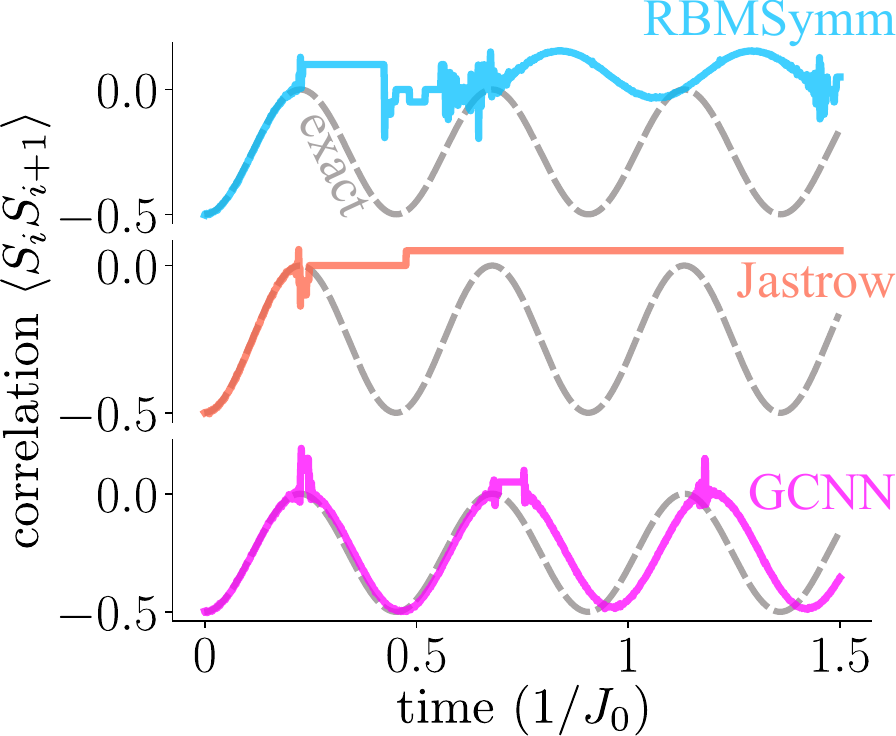}
\caption{Correlation dynamics obtained by neural network models available in NetKet: symmetrical RBM (RBMSymm), Jastrow ansatz, and group convolutional neural network (GCNN). All models show features of breakdown.} \label{fig:netket}
\end{figure}

The results are shown in Fig. \ref{fig:netket}. Importantly, to obtain an accurate estimation of correlation functions during these simulations, a very large number of Monte Carlo samples was needed; we used $50000$. Yet even with this large number of samples, the breakdown is observed. It should be noted that GCNN seems to lead to the most accurate results. However, it features the same instability around the correlation maxima as other models. The behaviour of the GCNN seems to be very similar to the results in the main text of the paper obtained with an RBM model and regularization as the inversion method  (see Fig. \ref{fig:breakdown} (c)), showing that after stationary correlation, dynamics can reappear. However, this reappearance happens at the wrong times compared with the exact results.

These results show that, in addition to varying RBM network width, the breakdown scenario is not limited to the RBM architecture.

\section{$\Psi - \log \Psi$ formulations} \label{sec:appendix_logpsi}
\begin{figure}[h!]
\centering
\includegraphics[scale=0.6]{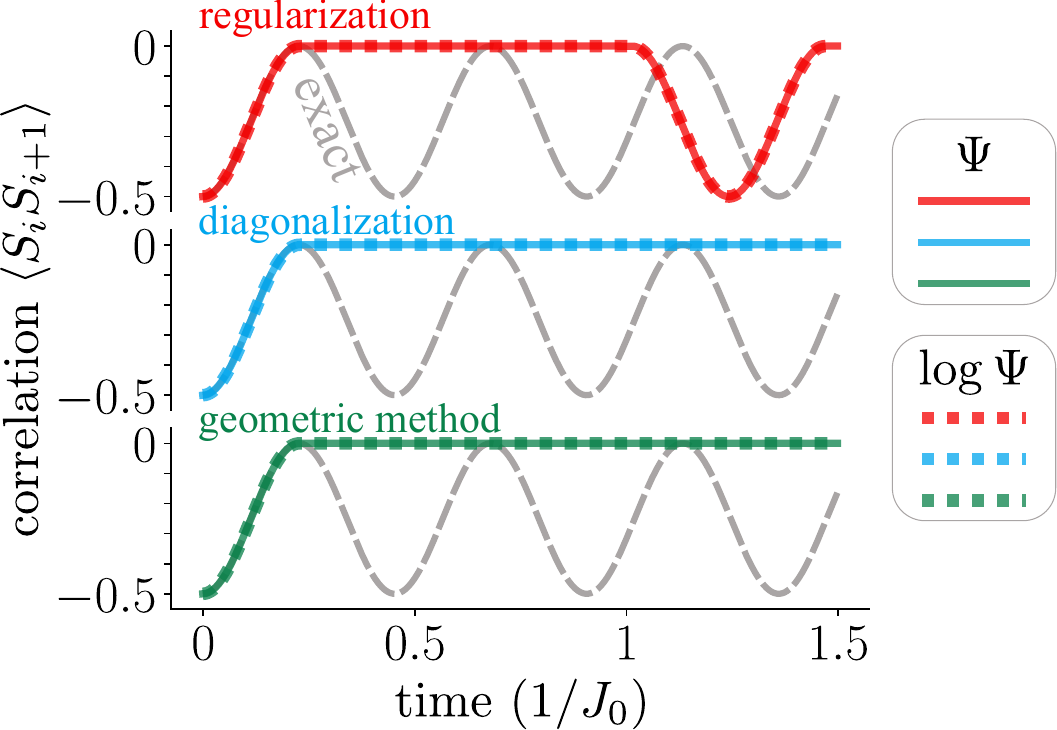}
\caption{Correlation dynamics for all three inversion methods used in this work, for the $\Psi$ (full lines) and the $\log \Psi$ (broken lines) formulations. The results are identical starting from the same initial state.} \label{fig:psi_logpsi}
\end{figure}

In this appendix, we compare two implementations for estimating quantum observables. The first implementation uses the wave function itself for the evaluation, as in the main text ($\Psi$-formulation). The second formulation deploys the logarithm of the wave function ($\log \Psi$-formulation), which is commonly done to avoid large-scale differences in the wave function values. Importantly, in the $\log \Psi$-formulation, a value of a single wave function element is never calculated, even in full summation. We find that both formulations yield the same dynamics, including the breakdown for $\Delta = -2$, as shown in Fig. \ref{fig:psi_logpsi}. Hence, we conclude that multiscale problems are not the origin of the observed instabilities.

\end{appendix}



\begin{thebibliography}{10}
\providecommand{\url}[1]{\texttt{#1}}
\providecommand{\urlprefix}{URL }
\expandafter\ifx\csname urlstyle\endcsname\relax
  \providecommand{\doi}[1]{doi:\discretionary{}{}{}#1}\else
  \providecommand{\doi}{doi:\discretionary{}{}{}\begingroup \urlstyle{rm}\Url}\fi
\providecommand{\eprint}[2][]{\url{#2}}

\bibitem{eisert_quantum_2015}
J.~Eisert, M.~Friesdorf and C.~Gogolin,
\newblock \emph{Quantum many-body systems out of equilibrium},
\newblock Nature Physics \textbf{11}(2), 124 (2015),
\newblock \doi{10.1038/nphys3215}.

\bibitem{lin_exact_1993}
H.~Lin, J.~Gubernatis, H.~Gould and J.~Tobochnik,
\newblock \emph{Exact {{Diagonalization Methods}} for {{Quantum Systems}}},
\newblock Computers in Physics \textbf{7}(4), 400 (1993),
\newblock \doi{10.1063/1.4823192}.

\bibitem{wu_variational_2023}
D.~Wu, R.~Rossi, F.~Vicentini, N.~Astrakhantsev, F.~Becca, X.~Cao, J.~Carrasquilla, F.~Ferrari, A.~Georges, M.~{Hibat-Allah}, M.~Imada, A.~M. L{\"a}uchli \emph{et~al.},
\newblock \emph{Variational {{Benchmarks}} for {{Quantum Many-Body Problems}}} (2023), \eprint{2302.04919}.

\bibitem{lu_universal_2020}
Y.~Lu and J.~Lu,
\newblock \emph{A {{Universal Approximation Theorem}} of {{Deep Neural Networks}} for {{Expressing Probability Distributions}}},
\newblock \doi{10.48550/arXiv.2004.08867} (2020), \eprint{2004.08867}.

\bibitem{rzadkowski_artificial_2022}
W.~Rzadkowski, M.~Lemeshko and J.~H. Mentink,
\newblock \emph{Artificial neural network states for nonadditive systems},
\newblock Physical Review B \textbf{106}(15), 155127 (2022),
\newblock \doi{10.1103/PhysRevB.106.155127}.

\bibitem{lange_architectures_2024}
H.~Lange, A.~Van De~Walle, A.~Abedinnia and A.~Bohrdt,
\newblock \emph{From architectures to applications: A review of neural quantum states},
\newblock Quantum Science and Technology \textbf{9}(4), 040501 (2024),
\newblock \doi{10.1088/2058-9565/ad7168}.

\bibitem{carleo_solving_2017}
G.~Carleo and M.~Troyer,
\newblock \emph{Solving the quantum many-body problem with artificial neural networks},
\newblock Science \textbf{355}(6325), 602 (2017),
\newblock \doi{10.1126/science.aag2302}.

\bibitem{schmitt_quantum_2020}
M.~Schmitt and M.~Heyl,
\newblock \emph{Quantum {{Many-Body Dynamics}} in {{Two Dimensions}} with {{Artificial Neural Networks}}},
\newblock Physical Review Letters \textbf{125}(10), 100503 (2020),
\newblock \doi{10.1103/PhysRevLett.125.100503}.

\bibitem{fabiani_supermagnonic_2021}
G.~Fabiani, M.~D. Bouman and J.~H. Mentink,
\newblock \emph{Supermagnonic {{Propagation}} in {{Two-Dimensional Antiferromagnets}}},
\newblock Physical Review Letters \textbf{127}(9), 097202 (2021),
\newblock \doi{10.1103/PhysRevLett.127.097202}.

\bibitem{zhang_paths_2025}
W.~Zhang, B.~Xing, X.~Xu and D.~Poletti,
\newblock \emph{Paths towards time evolution with larger neural-network quantum states},
\newblock Computer Physics Communications \textbf{312}, 109577 (2025),
\newblock \doi{10.1016/j.cpc.2025.109577}.

\bibitem{hofmann_role_2022}
D.~Hofmann, G.~Fabiani, J.~Mentink, G.~Carleo and M.~Sentef,
\newblock \emph{Role of stochastic noise and generalization error in the time propagation of neural-network quantum states},
\newblock SciPost Physics \textbf{12}(5), 165 (2022),
\newblock \doi{10.21468/SciPostPhys.12.5.165}.

\bibitem{czischek_quenches_2018}
S.~Czischek, M.~G{\"a}rttner and T.~Gasenzer,
\newblock \emph{Quenches near {{Ising}} quantum criticality as a challenge for artificial neural networks},
\newblock Physical Review B \textbf{98}(2), 024311 (2018),
\newblock \doi{10.1103/PhysRevB.98.024311}.

\bibitem{gravina_neural_2024}
L.~Gravina, V.~Savona and F.~Vicentini,
\newblock \emph{Neural {{Projected Quantum Dynamics}}: A systematic study},
\newblock \doi{10.48550/arXiv.2410.10720} (2024), \eprint{2410.10720}.

\bibitem{gutierrez_real_2022}
I.~L. Guti{\'e}rrez and C.~B. Mendl,
\newblock \emph{Real time evolution with neural-network quantum states},
\newblock Quantum \textbf{6}, 627 (2022),
\newblock \doi{10.22331/q-2022-01-20-627},
\newblock \eprint{1912.08831}.

\bibitem{lin_scaling_2022}
S.-H. Lin and F.~Pollmann,
\newblock \emph{Scaling of {{Neural}}-{{Network Quantum States}} for {{Time Evolution}}},
\newblock physica status solidi (b) \textbf{259}(5), 2100172 (2022),
\newblock \doi{10.1002/pssb.202100172}.

\bibitem{kramer_geometry_1981}
P.~Kramer and M.~Saraceno,
\newblock \emph{Geometry of the Time-Dependent Variational Principle in Quantum Mechanics},
\newblock No. 140 in Lecture Notes in Physics. Springer, Berlin,
\newblock ISBN 978-3-540-10579-4 978-0-387-10579-6 (1981).

\bibitem{kramer_review_2008}
P.~Kramer,
\newblock \emph{A review of the time-dependent variational principle},
\newblock Journal of Physics: Conference Series \textbf{99}, 012009 (2008),
\newblock \doi{10.1088/1742-6596/99/1/012009}.

\bibitem{haegeman_timedependent_2011}
J.~Haegeman, J.~I. Cirac, T.~J. Osborne, I.~Pi{\v z}orn, H.~Verschelde and F.~Verstraete,
\newblock \emph{Time-{{Dependent Variational Principle}} for {{Quantum Lattices}}},
\newblock Physical Review Letters \textbf{107}(7), 070601 (2011),
\newblock \doi{10.1103/PhysRevLett.107.070601}.

\bibitem{ido_timedependent_2015}
K.~Ido, T.~Ohgoe and M.~Imada,
\newblock \emph{Time-dependent many-variable variational {{Monte Carlo}} method for nonequilibrium strongly correlated electron systems},
\newblock Physical Review B \textbf{92}(24), 245106 (2015),
\newblock \doi{10.1103/PhysRevB.92.245106}.

\bibitem{hjorth-jensen_computational_2010}
M.~{Hjorth-Jensen},
\newblock \emph{Computational {{Physics}}, {{Lecture}} Notes},
\newblock University of Oslo,
\newblock \urlprefix\url{https://courses.physics.ucsd.edu/2017/Spring/physics142/Lectures/Lecture18/Hjorth-JensenLectures2010.pdf} (2010).

\bibitem{becca_quantum_2017}
F.~Becca and S.~Sorella,
\newblock \emph{Quantum {{Monte Carlo Approaches}} for {{Correlated Systems}}},
\newblock Cambridge University Press, 1 edn.,
\newblock ISBN 978-1-107-12993-1 978-1-316-41704-1,
\newblock \doi{10.1017/9781316417041} (2017).

\bibitem{donatella_dynamics_2023}
K.~Donatella, Z.~Denis, A.~Le~Boit{\'e} and C.~Ciuti,
\newblock \emph{Dynamics with autoregressive neural quantum states: {{Application}} to critical quench dynamics},
\newblock Physical Review A \textbf{108}(2), 022210 (2023),
\newblock \doi{10.1103/PhysRevA.108.022210}.

\bibitem{walle_manybody_2024}
A.~Van De~Walle, M.~Schmitt and A.~Bohrdt,
\newblock \emph{Many-body dynamics with explicitly time-dependent neural quantum states},
\newblock Machine Learning: Science and Technology \textbf{6}(4), 045011 (2025),
\newblock \doi{10.1088/2632-2153/ae0f39}.

\bibitem{sinibaldi_unbiasing_2023}
A.~Sinibaldi, C.~Giuliani, G.~Carleo and F.~Vicentini,
\newblock \emph{Unbiasing time-dependent {{Variational Monte Carlo}} by projected quantum evolution},
\newblock Quantum \textbf{7}, 1131 (2023),
\newblock \doi{10.22331/q-2023-10-10-1131},
\newblock \eprint{2305.14294}.

\bibitem{hackl_geometry_2020}
L.~Hackl, T.~Guaita, T.~Shi, J.~Haegeman, E.~Demler and I.~Cirac,
\newblock \emph{Geometry of variational methods: Dynamics of closed quantum systems},
\newblock SciPost Physics \textbf{9}(4), 048 (2020),
\newblock \doi{10.21468/SciPostPhys.9.4.048}.

\bibitem{fabiani_quantum_2022}
G.~Fabiani,
\newblock \emph{Quantum Dynamics of {{2D}} Antiferromagnets: Predictions from Theory and Machine Learning},
\newblock Ph.D. thesis, Radboud University,
\newblock \doi{https://hdl.handle.net/2066/250503} (2022).

\bibitem{king_computational_2024}
A.~D. King, A.~Nocera, M.~M. Rams, J.~Dziarmaga, R.~Wiersema, W.~Bernoudy, J.~Raymond, N.~Kaushal, N.~Heinsdorf, R.~Harris, K.~Boothby, F.~Altomare \emph{et~al.},
\newblock \emph{Computational supremacy in quantum simulation},
\newblock \doi{10.48550/arXiv.2403.00910} (2024), \eprint{2403.00910}.

\bibitem{sinibaldi_timedependent_2024}
A.~Sinibaldi, D.~Hendry, F.~Vicentini and G.~Carleo,
\newblock \emph{Time-dependent {{Neural Galerkin Method}} for {{Quantum Dynamics}}},
\newblock \doi{10.48550/arXiv.2412.11778} (2024), \eprint{2412.11778}.

\bibitem{kelly_adaptive_2022}
C.~Kelly and G.~J. Lord,
\newblock \emph{Adaptive {{Euler}} methods for stochastic systems with non-globally {{Lipschitz}} coefficients},
\newblock Numerical Algorithms \textbf{89}(2), 721 (2022),
\newblock \doi{10.1007/s11075-021-01131-8}.

\bibitem{dash_efficiency_2025}
S.~Dash, L.~Gravina, F.~Vicentini, M.~Ferrero and A.~Georges,
\newblock \emph{Efficiency of neural quantum states in light of the quantum geometric tensor},
\newblock Communications Physics \textbf{8}(1), 92 (2025),
\newblock \doi{10.1038/s42005-025-02005-4}.

\bibitem{zhao_magnon_2004}
J.~Zhao, A.~V. Bragas, D.~J. Lockwood and R.~Merlin,
\newblock \emph{Magnon {{Squeezing}} in an {{Antiferromagnet}}: {{Reducing}} the {{Spin Noise}} below the {{Standard Quantum Limit}}},
\newblock Physical Review Letters \textbf{93}(10), 107203 (2004),
\newblock \doi{10.1103/PhysRevLett.93.107203}.

\bibitem{bossini_macrospin_2016}
D.~Bossini, S.~Dal~Conte, Y.~Hashimoto, A.~Secchi, R.~V. Pisarev, {\relax Th}.~Rasing, G.~Cerullo and A.~V. Kimel,
\newblock \emph{Macrospin dynamics in antiferromagnets triggered by sub-20 femtosecond injection of nanomagnons},
\newblock Nature Communications \textbf{7}(1), 10645 (2016),
\newblock \doi{10.1038/ncomms10645}.

\bibitem{bossini_laserdriven_2019}
D.~Bossini, S.~Dal~Conte, G.~Cerullo, O.~Gomonay, R.~V. Pisarev, M.~Borovsak, D.~Mihailovic, J.~Sinova, J.~H. Mentink, {\relax Th}.~Rasing and A.~V. Kimel,
\newblock \emph{Laser-driven quantum magnonics and terahertz dynamics of the order parameter in antiferromagnets},
\newblock Physical Review B \textbf{100}(2), 024428 (2019),
\newblock \doi{10.1103/PhysRevB.100.024428}.

\bibitem{formisano_coherent_2024}
F.~Formisano, T.~T. Gareev, D.~I. Khusyainov, A.~E. Fedianin, R.~M. Dubrovin, P.~P. Syrnikov, D.~Afanasiev, R.~V. Pisarev, A.~M. Kalashnikova, J.~H. Mentink and A.~V. Kimel,
\newblock \emph{Coherent {{THz}} spin dynamics in antiferromagnets beyond the approximation of the {{N{\'e}el}} vector},
\newblock APL Materials \textbf{12}(1), 011105 (2024),
\newblock \doi{10.1063/5.0180888}.

\bibitem{choo_symmetries_2018}
K.~Choo, G.~Carleo, N.~Regnault and T.~Neupert,
\newblock \emph{Symmetries and {{Many-Body Excitations}} with {{Neural-Network Quantum States}}},
\newblock Physical Review Letters \textbf{121}(16), 167204 (2018),
\newblock \doi{10.1103/PhysRevLett.121.167204}.

\bibitem{carrasquilla_machine_2020}
J.~Carrasquilla,
\newblock \emph{Machine learning for quantum matter},
\newblock Advances in Physics: X \textbf{5}(1), 1797528 (2020),
\newblock \doi{10.1080/23746149.2020.1797528}.

\bibitem{fabiani_investigating_2019}
G.~Fabiani and J.~Mentink,
\newblock \emph{Investigating ultrafast quantum magnetism with machine learning},
\newblock SciPost Physics \textbf{7}(1), 004 (2019),
\newblock \doi{10.21468/SciPostPhys.7.1.004}.

\bibitem{fabiani_ultrafast_2022}
G.~Fabiani and J.~H. Mentink,
\newblock \emph{Ultrafast dynamics of entanglement in {{Heisenberg}} antiferromagnets},
\newblock Physical Review B \textbf{105}(9), 094438 (2022),
\newblock \doi{10.1103/PhysRevB.105.094438}.

\bibitem{penrose_generalized_1955}
R.~Penrose,
\newblock \emph{A generalized inverse for matrices},
\newblock Mathematical Proceedings of the Cambridge Philosophical Society \textbf{51}(3), 406 (1955),
\newblock \doi{10.1017/S0305004100030401}.

\bibitem{novak_numerical_2022}
K.~Novak,
\newblock \emph{Numerical {{Methods}} for {{Scientific Computing}}},
\newblock Equal Share Press, second edition edn.,
\newblock ISBN 979-8-9854218-0-4 (2022).

\bibitem{virtanen_scipy_2020}
P.~Virtanen, R.~Gommers, T.~E. Oliphant, M.~Haberland, T.~Reddy, D.~Cournapeau, E.~Burovski, P.~Peterson, W.~Weckesser, J.~Bright, S.~J. Van Der~Walt, M.~Brett \emph{et~al.},
\newblock \emph{{{SciPy}} 1.0: Fundamental algorithms for scientific computing in {{Python}}},
\newblock Nature Methods \textbf{17}(3), 261 (2020),
\newblock \doi{10.1038/s41592-019-0686-2}.

\bibitem{ultrafast-code_ultrafastcode_2024}
G.~Fabiani and J.~H. Mentink,
\newblock \emph{Ultrafast},
\newblock \urlprefix\url{https://github.com/ultrafast-code/ULTRAFAST} (2024).

\bibitem{marshall_antiferromagnetism_1955}
W.~Marshall,
\newblock \emph{Antiferromagnetism},
\newblock Proceedings of the Royal Society of London. Series A, Mathematical and Physical Sciences \textbf{232}(1188), 48 (1955),
\newblock \eprint{99682}.

\bibitem{ledinauskas_universal_2025}
E.~Ledinauskas and E.~Anisimovas,
\newblock \emph{Universal performance gap of neural quantum states applied to the {{Hofstadter-Bose-Hubbard}} model},
\newblock SciPost Physics \textbf{18}(1), 011 (2025),
\newblock \doi{10.21468/SciPostPhys.18.1.011}.

\bibitem{vicentini_netket_2022}
F.~Vicentini, D.~Hofmann, A.~Szab{\'o}, D.~Wu, C.~Roth, C.~Giuliani, G.~Pescia, J.~Nys, V.~{Vargas-Calder{\'o}n}, N.~Astrakhantsev and G.~Carleo,
\newblock \emph{{{NetKet}} 3: {{Machine Learning Toolbox}} for {{Many-Body Quantum Systems}}},
\newblock SciPost Physics Codebases p.~7 (2022),
\newblock \doi{10.21468/SciPostPhysCodeb.7}.

\bibitem{golub_matrix_2013}
G.~H. Golub and C.~F. Van~Loan,
\newblock \emph{Matrix Computations},
\newblock Johns {{Hopkins}} Studies in the Mathematical Sciences. The Johns Hopkins University Press, Baltimore, fourth edition edn.,
\newblock ISBN 978-1-4214-0794-4 (2013).

\bibitem{harris_array_2020}
C.~R. Harris, K.~J. Millman, S.~J. Van Der~Walt, R.~Gommers, P.~Virtanen, D.~Cournapeau, E.~Wieser, J.~Taylor, S.~Berg, N.~J. Smith, R.~Kern, M.~Picus \emph{et~al.},
\newblock \emph{Array programming with {{NumPy}}},
\newblock Nature \textbf{585}(7825), 357 (2020),
\newblock \doi{10.1038/s41586-020-2649-2}.

\bibitem{paige_solution_1975}
C.~C. Paige and M.~A. Saunders,
\newblock \emph{Solution of {{Sparse Indefinite Systems}} of {{Linear Equations}}},
\newblock SIAM Journal on Numerical Analysis \textbf{12}(4), 617 (1975),
\newblock \doi{10.1137/0712047}.

\bibitem{cash_variable_1990}
J.~R. Cash and A.~H. Karp,
\newblock \emph{A variable order {{Runge-Kutta}} method for initial value problems with rapidly varying right-hand sides},
\newblock ACM Transactions on Mathematical Software \textbf{16}(3), 201 (1990),
\newblock \doi{10.1145/79505.79507}.

\bibitem{roth_group_2021}
C.~Roth and A.~H. MacDonald,
\newblock \emph{Group {{Convolutional Neural Networks Improve Quantum State Accuracy}}},
\newblock \doi{10.48550/arXiv.2104.05085} (2021), \eprint{2104.05085}.

\bibitem{klambauer_selfnormalizing_2017}
G.~Klambauer, T.~Unterthiner, A.~Mayr and S.~Hochreiter,
\newblock \emph{Self-{{Normalizing Neural Networks}}},
\newblock \doi{10.48550/arXiv.1706.02515} (2017), \eprint{1706.02515}.

\end{thebibliography}
\end{document}